\documentclass[prd,aps,floats,floatfix,eqsecnum,nofootinbib,12pt]{revtex4-1}
\usepackage{verbatim,graphicx,amssymb,amsbsy,bm,amsmath,rotating,epsfig}
\usepackage{psfrag}
\usepackage{hyperref}
\usepackage{fontenc}
\usepackage[latin1]{inputenc}
\usepackage{pstricks,pst-node,pst-text,pst-3d}
\usepackage{textcomp}
\newcommand{\be}{\begin{equation}}
\newcommand{\ee}{\end{equation}}
\newcommand{\bea}{\begin{eqnarray}}
\newcommand{\eea}{\end{eqnarray}}

\begin{document}

\title{New Quantum Structure of the Space-Time}
\author{Norma G. SANCHEZ} 
\affiliation{ LERMA CNRS UMR 8112 Observatoire de Paris PSL \\
Research University, Sorbonne Universit\'e UPMC Paris VI,\\ 
61, Avenue de l'Observatoire, 75014 Paris, France}
\date{\today }

\begin{abstract}
{\bf Abstract}: Starting from {\it quantum theory} 
(instead of general relativity) to approach quantum gravity  
within a minimal setting allows us here to describe the quantum 
space-time structure and the quantum light cone.
From the classical-quantum duality and quantum harmonic
oscillator $(X, P)$ variables in global phase space 
we promote the space-time coordinates to quantum non-commuting operators.
The phase space instanton $(X, P = iT)$ describes the  
hyperbolic quantum space-time structure and generates the {\it quantum light cone}.
The classical Minkowski space-time null generators $X = \pm T$ {\it dissapear} 
at the quantum level due to the relevant quantum
$[X,T]$ conmutator which is {\it always} non-zero. A {\it new} 
quantum Planck scale vacuum region emerges.
We describe the quantum Rindler and 
quantum Schwarzshild-Kruskal space-time structures. 
The horizons and the $r= 0$ space-time singularity are
 {\it quantum mechanically erased}. 
The four Kruskal regions merge inside a single quantum Planck scale "world". 
The quantum space-time structure consists of {\it hyperbolic discrete levels} 
of odd numbers $(X^2 - T^2)_n = (2n + 1)$ (in Planck units ), $n= 0, 1, 2...$. 
.($X_n, T_n$) and the {\it mass levels} being $\sqrt {(2n +1)}$. 
A coherent picture emerges:
large $n$ levels are semiclassical tending towards a 
classical continuum space-time. Low $n$ are quantum, the lowest mode ($n = 0$) 
being the Planck scale. Two dual $(\pm)$ branches are present in the local
variables ($\sqrt{2n+1}\pm \sqrt{2n}$) reflecting the duality of the 
large and small $n$ behaviours and covering the {\it whole} mass spectrum:
from the largest astrophysical objects in branch (+)
to the quantum elementary particles in branch (-) passing by the Planck mass. 
Black holes belong to both branches (+) and (-).\\ 
Norma.Sanchez@obspm.fr,\qquad \;
\url{https://chalonge-devega.fr/sanchez}
\end{abstract}
\maketitle
\tableofcontents

\section{Introduction and Results}

Recently, we extended  the known classical-quantum duality to include 
gravity and the Planck scale domain ref [1].
This led us to introduce more complete variables $O_{QG}$ fully taking into 
account all domains, classical and quantum gravity domains and their duality properties, 
passing by the Planck scale and the elementary particle range. 

One of the results of such study is the classical-quantum duality of the 
 Schwarzschild-Kruskal space-time.

\medskip

In this paper we go further in exploring the space-time structure with quantum theory 
and the Planck scale domain. The classical-quantum duality including
gravity and the QG variables are a  key insight in this study.
From the usual gravity (G) variables and quantum (Q) variables ($O_G, O_Q$), we introduced QG variables 
$O_{QG}$ which in units of the corresponding Planck scale magnitude $o_P$ simply read:
\be
O = \frac{1}{2}\;(\; x + \frac{1}{x}\;),\qquad  O\equiv \frac{O_{QG}}{o_P},
\qquad x \equiv \frac{O_G}{o_P} = \frac{o_P}{O_Q} 
\ee
The $QG$ variables automatically  are endowed with the symmetry 
\be
O(1/x ) = O(x)\; \mbox{and satisfy} \; O(x=1) = 1 \; \mbox{at the Planck scale}. 
\ee
QG variables are {\it complete} or {\it global}. 
{\it Two} values $ x_{\pm}$ of the usual variables $O_G$ or $O_Q$
are necessary for each variable {QG}. The $(+)$ and $(-)$ branches precisely 
correspond to the two different and dual ways of reaching the Planck scale: 
from the quantum elementary particle side ($0\leq x \leq 1$) and 
from the classical/semiclassical gravity side ($1 \leq  x \leq \infty$). 
There is thus a classical-quantum {\bf duality} between 
the two domains. The gravity domain is dual  
(in the precise sense of the wave-particle duality) of the quantum elementary particle domain through the Planck scale: 
\be
O_{G}=o_{P}^2 \; O_Q^{-1}
\ee
Each of the sides of the duality Eq (1.3) accounts for only one domain: 
$Q$ or $G$  but not for both domains together. QG variables account for 
both of them, they contain the duality Eq.(1.3) and satisfy the {\it QG duality} 
Eq (1.2).

As the wave-particle duality, QG duality is general, it does not relate to the number of dimensions nor to any other condition.

\medskip

In particular, length and time, basic QG variables $(X,T)$
in their respective Planck units are:
\begin{equation}
X = \frac{1}{2} \;(x + \frac{1}{x}),   
\quad X \;\equiv \;\frac{L_{QG}}{l_P}
\end{equation} 
\be
T =  \frac{1}{2}\;(\;t-\frac{1}{t}\;), \quad T \;\equiv \;\frac{T_{QG}}{t_P}
\ee
$l_P$ and $t_{P}$ being the Planck length and time respectively. 
The usual variables stand here in lowercase letters. 
QG mass and momemtum variables are similar to $(X,T)$:
\be 
P =  \frac{1}{2}\;( p - \frac{1}{p}\;), \quad P \;\equiv \;\frac{P_{QG}}{p_P}
\ee
\be
M  = \frac{1}{2}\;( x + \frac{1}{x} ),   
 \quad M \;\equiv \;\frac{M_{QG}}{m_P},  \quad x \equiv \frac{m}{m_P}
\ee
These are pure numbers (in Planck units), the space can be parametrized by lengths or masses. 
$m$ is the usual mass variable and $m_P$ is the Planck mass.

The complete manifold of QG variables requires several "patches" 
or analytic extensions to cover the full sets $X \geq 1$ or $X \leq 1$:
\begin{equation}
x_{\pm} =  X \pm\sqrt{X^2 - 1}, \quad X \geq 1, \qquad
x_{\pm } =  X \pm \sqrt{1 - X^2}, \quad X \leq 1
\end{equation}
The  two $(X \geq 1)$, $(X \leq 1)$ domains being the classical and quantum domains 
respectively with their two $(\pm)$ branches each, and when $ 
x_{+} = x_{-}$:  $X = 1,\;  x_{\pm} = 1$,  (the Planck scale).
The QG variables $(X,T)$ satisfy:  
\be
X(x) = X(1/x),  \quad X(-x) = -X (x), \quad X(1) = 1
\ee
\be
T(t) = -T(1/t) ,  \quad  T(-t) = -T (t) , \quad T(1) = 0
\ee

QG variables can be also considered in phase-space ($X, P)$
with their full global analytic extension
as we describe in this paper. Comparison of the QG variables 
with the complete Q-variables of the harmonic oscillator is enlighting,
as we do in section II here.

\medskip

In {\bf this paper}, by promoting the QG variables $(X,T)$ to quantum non-commutative
coordinates, further insight into the quantum space-time structure is obtained and 
{\it new} results do appear.  

As already mentionned,  we take quantum theory as  
 the guide, and start by the " prototype case": the harmonic oscillator.

We find the quantum structure of the space-time
arising from the relevant non-zero space-time commutator $[X, T]$, 
or non-zero quantum uncertainty $ \Delta X \Delta T$ by considering {\it quantum} 
coordinates $(X,T)$.
All other commutators are zero. The remaining transverse spatial coordinates $X_{\bot}$ 
have all their commutators zero.

\medskip

The {\bf results of this paper} are the following:

\begin{itemize}
\item {We find the {\it quantum light cone}: It is generated by the quantum Planck hyperbolae 
$ X^2 - T^2 = \pm [X, T]$ due to the quantum uncertainty $[X,T]= 1$.
They replace the classical light cone generators $ X = \pm T $ which are
{\it quantum mechanically erased}. Inside the Planck hyperbolae there is a enterely 
{\it new quantum region} within the Planck scale and below which is purely quantum vacuum or zero-point energy.}
\item{In higher dimensions, the quantum commuting coordinates $(X,T)$ 
and the transverse non-commuting spatial coordinates $X_{\bot j}$
generate the quantum two-sheet hyperboloid 
$ X^2 - T^2 + X_{\bot j}\; X_{\bot}^j = \pm 1$, \;
$ {\bot} j = 2, 3, ... (D-2)$, $D$ being the total space-time dimensions, 
 $D=4$ in particular in the cases considered here.}
\item{To quantize Minkowski space-time, we just consider quantum 
non-commutative coordinates $(X,T)$ with the usual (non deformed) canonical quantum commutator 
$[X,T] = 1$, ($1$ is here $l_P^2$), and all other commutators zero. In light-cone coordinates 
$$ 
U = \; \frac{1}{\sqrt{2}} \;(X - T), 
\qquad V = \; \frac{1}{\sqrt{2}} \;(X + T),
$$ 
the quadratic form  (symmetric order of operators) 
$s^2  = [UV + VU] = X^2 - T^2 = (2 VU + 1)$
determines the relevant part of the quantum distance. Upon identification $T = -iP$, 
the quantum coordinates $(U, V)$ for hyperbolic space-time are precisely the ($a, a^+
$) operators for euclidean phase space (the phase space {\it instanton}) and as a 
consequence $VU$ is the Number operator. 
The expectation value $ (s^2)_n = (2n + 1)$ has a minimal non zero value: $(s^2)_
{n=0} = 1$ which is the zero point energy or Planck scale vacuum. 
Consistently, in quantum space-time: $$ (T^2 - X^2) - 1 \geq 0 : \;\mbox {{\it 
timelike}} $$
$$ (X^2 - T^2) - 1 \geq 0:  \;\mbox {{\it spacelike}} $$
$$ (T^2 - X^2) - 1 = 0, \; \mbox {{\it null}\;: the {\it"quantum light-cone"}}.$$
This shows that only outside the null hyperbolae, that is outside the Planck 
scale vacuum region, such notions as distance, and timelike and spacelike signatures,
 can be defined, Section III and Figs 3, 4.}
\item{Here we quantized the $(X, T)$ dimensions  which are relevant 
to the light-cone space-time structure, as this
is the case for the Rindler, Schwarzschild - Kruskal and other manifolds.
The remaining spatial transverse dimensions  $ X_{\bot} $ are considered here 
as commuting coordinates. For instance, 
in Minkowski space-time:
\be
s^2 = (X^2 - T^2 + X_{\bot j} X_{\bot }^j ), \qquad {\bot j} = 2,3, ...(D-2).
\ee
\be
[X_{\bot j},X] = 0 = [X_{\bot j},T], \qquad [X_{\bot i}, X_{\bot j}] = 0, 
\qquad [P_{\bot i}, P_{\bot j}] = 0
\ee
for all ${\bot} i,j = 1, ....., (D-2)$, \; $D$ being the total space-time dimensions.

\medskip

This corresponds to quantize the two-dimensional surface $(X,T)$ relevant 
for the light-cone structure, 
leaving the transverse spatial dimensions ${\bot}$ essentially unquantized 
(although they have zero commutators
they could fluctuate). This is enough for considering the 
new features arising in the  {\it quantum light cone} and in the 
quantum Rindler and the quantum Schwarzschild-Kruskal space-time structures, 
for which as is known, the relevant classical structures are in the $(X,T)$ dimensions 
and not in the transverse spatial ${\bot}$ ones. Quantum manifolds 
where the transverse space $X_{\bot}$ coordinates are 
non-commuting will be considered elsewhere.} 
\item{We find the quantum Rindler and the quantum Schwarzschild-Kruskal
space-time structures. At the quantum level, the classical null horizons 
$X = \pm \; T$ 
are {\it erased}, and the $r=0$ classical singularity {\it dissapears}. 
The space-time structure turns out to be {\it discretized}
in {\it quantum} hyperbolic levels $X_n^2 - T_n^2 = \pm (2n+1), \; n = 0,1,2...$.
For large $n$ the space-time becomes classical and continuum. 
Moreover, the classical singular $r= 0$ hyperbolae are quantum mechanically {\it 
excluded}, they do {\it not} belong to any of the quantum allowed levels.}
\item{We find the mass quantization for {\it all} masses. 
The quantum mass levels are associated to the quantum space-time structure. 
The global mass levels are $M_n = m_P \sqrt {2n+ 1}$ \; for all $n= 0, 1, 2, ...$. 
{\it Two}  dual
branches $m_{n\pm} = m_P \; [\sqrt{2n +1 } \pm \sqrt{2n}\;] $ do appear 
for the usual mass variables, covering the {\it whole mass range}: from the 
Planck mass ($n=0$) till the largest astronomical masses: 
gravity branch (+), and from zero mass $(n = \infty)$ till near the Planck mass: 
elementary particle branch (-). For large $n$, masses {\it increase} as $m_P (2\sqrt 
{2n})$ in branch (+) while they {\it decrease} as $m_P /(2\sqrt{2n})$ in branch (-). 
For very large $n$ the spectrum becames continuum. Black holes belong to 
both branches (+) and (-); quantum strings have similar mass quantization.
In the conclusions we comment on these aspects.} 
\item {The end of black hole evaporation is not the subject of this paper  
but our results here have implications for it. Black hole ends its evaporation 
in branch (-). We know from refs [2],[7],[8] that it decays like a quantum heavy 
particle in pure (non mixed) states. 
In its last phase (mass smaller than the Planck mass $m_P$), the state 
{\it is not anymore like a black hole but like a heavy particle}. We discuss more on 
it in the conclusions.}
\end{itemize}
This paper is organized as follows: In
Section II we describe quantum space-time as a quantum harmonic oscillator and 
its classical-quantum duality properties. 
 In Section III we describe the quantum Rindler space-time and its structure.
Section IV deals with the quantum Schwarzschild-Kruskal space-time 
and its properties. In Section V we treat the quantized whole mass spectrum. 
In Section VI we present our remarks and conclusions.

\section {Quantum Space-Time as a Harmonic Oscillator}

 Comparison of the QG variables to the harmonic oscillator variables 
is enlighting. Let us first consider the complete variables not 
yet promoted to quantum non-conmuting operators. 
The oscillator complete variables $(X,P)$  
containing both the classical and quantum components are:
$$
X_Q = \frac{l}{2}\;(\frac{l}{\hbar}\; p + \frac{\hbar} {l\; p}), \qquad
P_Q =  \frac{\hbar}{2 l}\;(\frac{l}{\hbar} p - \frac{\hbar}{l\; p}\;), 
\qquad {l} = \frac{2\pi}{\omega} 
$$
being $ l$ the length of the oscillator, (also expressed as $\sqrt{\hbar m /\omega }$.

Or, in dimensionless variables:
$$
X = \frac{1}{2}\;(p +\frac{1}{p}), \qquad P = \;\frac{1}{2} \;(p - \frac{1}{p}),
\quad p \equiv \frac{l}{\hbar} p, \qquad X \equiv \;\frac{X_Q}{l}, 
\qquad P \;  \equiv \frac{l}{\hbar}\;P_Q
$$
There are two branches $p_{\pm}$ for each variable $X$ or $P$ and 
the two domains $X \geq 1 $ and $ X \leq 1$ are dual of each other, classical and quantum 
ones respectively: 

\begin{itemize}

\item{Classical: $X^2 >> 1$; 
Transition: $X^2 \simeq 1$, $ p_{+} = p_{-} = 1$; Quantum: $X^2 \leq 1$}.

\end{itemize}

Or, in terms of the star variables  $ p = \exp  p* $ :
$$ 
X =  \cosh p*, \quad  P = \sinh p*,\quad  X^2 - P^2 = 1.
$$ 
The value $ {l} = 1$,  ie $\hbar = m\omega $, (quantum action and classical momentum equal) 
is here the analogous of the Planck scale for QG, ie the transition from the classical  
$(m\omega >> \hbar)$ regime to the quantum $(m\omega  << \hbar)$ regime.
The hyperbolae $X^2 - P^2 = 1$, or fully dimensional 
$
X_Q^2 / l^2 - (l^2 P_Q^2)/\hbar^2 = 1,
$
are the transition "boundaries" between the classical or semiclassical and
the quantum regions
in the complete analytic extension of the ($X, P$) manifold. 
This is a {\it hyperbolic} phase space structure. Fig. 1 displays the four regions: 

\begin{itemize}

\item {Right and left exterior regions to the hyperbola $X^2 - P^2 = \pm 1$
, $\left| X \right| \geq P$ and $\left| X\right| \leq |P|$
are classical: $ X >> 1 $: $m \omega >> \hbar$}

\item {The hyperbolae $ X^2 - P^2 = \pm 1$ are the transition boundaries
$l \simeq 1: m \omega \simeq \hbar$. They 
separate the classical from the semiclassical and quantum regions.}

\item {"Future" and "past" interior regions $P > 0$ and $P < 0$ 
are quantum: $ X << 1 $: $m \omega  << \hbar$}

\end {itemize}

\begin{figure}
\includegraphics[height=15.cm,width=16.cm]{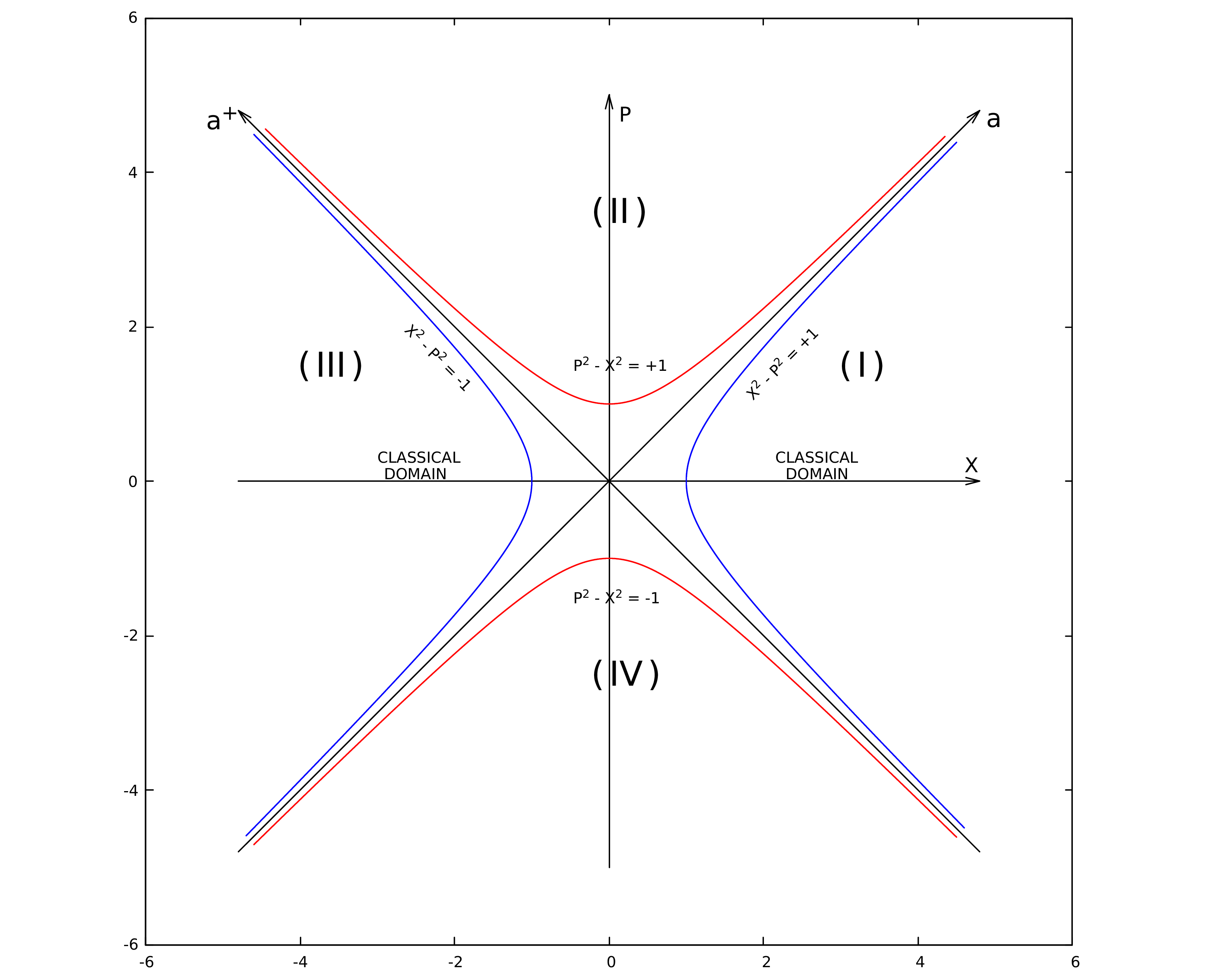}
\caption{ The complete analytic extension of the $(X, P)$ 
quantum harmonic oscillator variables and its classical 
and quantum domains: {\bf Hyperbolic phase space}. 
The $(a, a^+)$ operators are like light-cone coordinates. 
The {\it instanton} $P \rightarrow i P$, is the usual 
(elliptic) phase space with (dimensionless) hamiltonian 
$(X^2 + P^2) = 2 H$, (in units of the typical oscillator length).}
\label{fig1}
\end{figure}

Extension of $P$ to be purely imaginary: $P \rightarrow i P$, $p*\rightarrow ip*$,
(ie {\it instanton})
goes from the hyperbolic to the {\it elliptic} phase space structure  with the Hamiltonian  
$ H = (X^2 + P^2)/2 $, or in the dimensionfull variables:
$$
H_Q = \frac{\omega\hbar}{2} \; (\frac{X_Q^2}{l^2} + \frac{l^2 P_Q^2}{\hbar^2}), 
\qquad H \equiv \frac{H_Q} {\omega \hbar}
$$

\bigskip
 
By promoting $(X, P)$ to be quantum operators,
in terms of the $(a, a^+)$ representation yields:
\be
X = \; \frac {1}{\sqrt{2}} \; (a^+ + a), \qquad  
P = \; \frac {i}{\sqrt{2}} \; (a^+ - a),  \qquad [a, a^+] = 1,   
\ee
\be
2 H = (X^2 + P^2) = (aa^+ + a^+ a) = 2\;( a^+ a + \frac{1}{2}), \qquad
(X^2 - P^2) = (a^2 + a^{+ 2}) 
\ee
$$
[2H, P]  = i X, \qquad \;  [2H, X]  = -i P, \qquad \; [X, P]  = i, 
$$
\be \mbox {with the quantum levels} \; \; \epsilon _n = ( n  + \frac{1}{2} ), 
\; \; n = 0, 1, ...\ee
These are the dimensionless levels, (otherwise they are multiplied by $\omega \hbar$).

\medskip

The $(a, a^+)$ operators are the {\it light-cone} type quantum coordinates of 
the phase space $(X, P)$: 
\be a = \; \frac {1}{\sqrt{2}} \;(X + i P), 
\qquad a^+ = \; \frac {1}{\sqrt{2}} \;(X - i P)\ee
The temporal variable $T$ in the space-time configuration $(X, T)$ 
is like the (imaginary) momentum in phase space $(X, P)$: The
identification $P = iT$ in  Eqs (2.1)-(2.3) yields:
\be
X = \; \frac {1}{\sqrt{2}} \; (a^+ + a), \qquad  
T = \; \frac {1}{\sqrt{2}} \; (a^+ - a),  \qquad [a, a^+] = 1,   
\ee
\be
2H = (X^2 - T^2) = 2\; (a^+ a + \frac{1}{2}), \; \qquad 
(X^2 + T^2) = (a^2 + a^{+ 2}),
\ee
\be
[2H, T]  = X, \qquad \;  [2H, X]  = T,  \qquad \; [X, T] = 1,
\ee
$ a^+\;a = N $ being the number operator.

\medskip

Regions I, II, III, IV, corresponding to the 
exterior and interior regions to the 
hyperbolae $ X^2  \geq (T^2 \pm 1) $, $ X^2 \leq(T^2 \pm 1) $ respectively, 
are covered by patches similar to the (space-like) Eqs.(2.5)-(2.7). 
$ X $ and $T$ are interchanged in the time-like regions, similar to the global 
hyperbolic structure Fig 1.

\medskip
  
Given the quantum hyperbolic space-time structure above described ,
we can think then the quantum space-time coordinates $(X,T)$ as quantum harmonic 
oscillator coordinates $(X, T = iP)$, including quantum space-time fluctuations 
with length and mass in the Planck scale domain and quantized levels, as described by 
Eqs (2.5)-(2.7):
$$
0 \leq \; l  \; \leq l_P, \quad \epsilon _n = ( n + \frac{1}{2} ), \; \; n = 1, 
2, ... , 
\qquad \omega = 2 \pi /l
$$
Expectation values of Eqs (2.6) yield 
\be
(X^2 - T^2)_n \; = \;  2 \;( n + \frac{1}{2} )
\ee

The quantum algebra Eqs (2.5)-(2.7) describe the basic {\it quantum space-time} structure.   
\begin {itemize}
\item {When $[X, T] = 0$, they  yield the characteristic lines 
and light cones generators $ X= \pm T $ of the classical space-time structure and its causal 
domains, (Fig.2).}
\item{At the quantum level, the corresponding characteristic lines and light cone 
generators Eqs (2.6)-(2.8) are {\it bent} by the relevant $[X,T]$ commutator, they
do {\it not cross} at $X = \pm T = 0$ but are separated by the {\it quantum hyperbolic 
region} $2\epsilon_0$ due to the zero point energy (or quantum space-time width) 
$\epsilon_0 = (1/2) [X,T]$:
\be
(X^2 - T^2) = \pm [X,T] =\pm 1, \; 1 = 2 \epsilon_0 ,\; (n=0):
\quad \mbox{the quantum light cone}
\ee
$$ [X,T] = 0:  \qquad X = \pm T:  \qquad \mbox{the classical light cone}.$$}
\item{The hyperbolae Eq.(2.9) are the {\it quantum light cone}. 
They  quantum generalize the 
classical light cone $X = \pm T$ generators when $ [X, T]= 0 $.  The classical generators are   
the asymptotes for $T \rightarrow \pm \infty$. Quantum mechanically, $ X$ {\it is always different} from $\pm T$  since $[X, T]$ is always  different from zero. Figs 2-3 
illustrate these properties:
The well known classical (non quantum) light cone generators and the new {\it quantum 
light cone} (quantum Planck hyperbolae) due to the  $ 2 \epsilon_0$ zero-point energy.}
\item{Quantum fluctuations and the quantum generated thickness make the space-time 
structure {\it spread}, and its signature or causal structure is {\it quantum 
mechanically} modified, entangled, or {\it erased} in the 
quantum Planck scale region.}
\end{itemize}

\begin{figure}
\includegraphics[height=12.cm,width=13.cm]{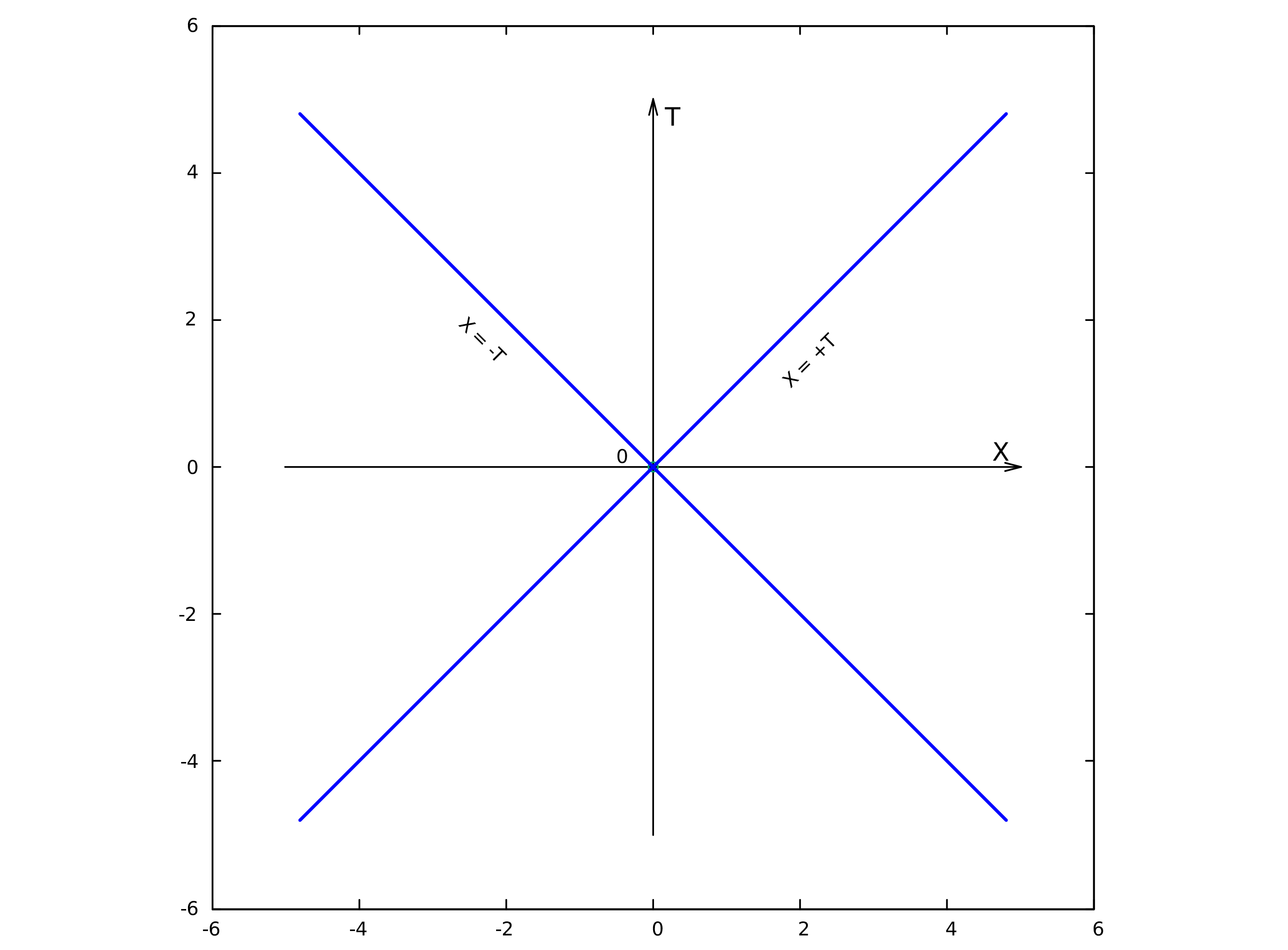}
\caption{\bf{The classical light cone.}}
\label{fig2a}
\end{figure}
\begin{figure}
\includegraphics[height=14.cm,width=16.cm]{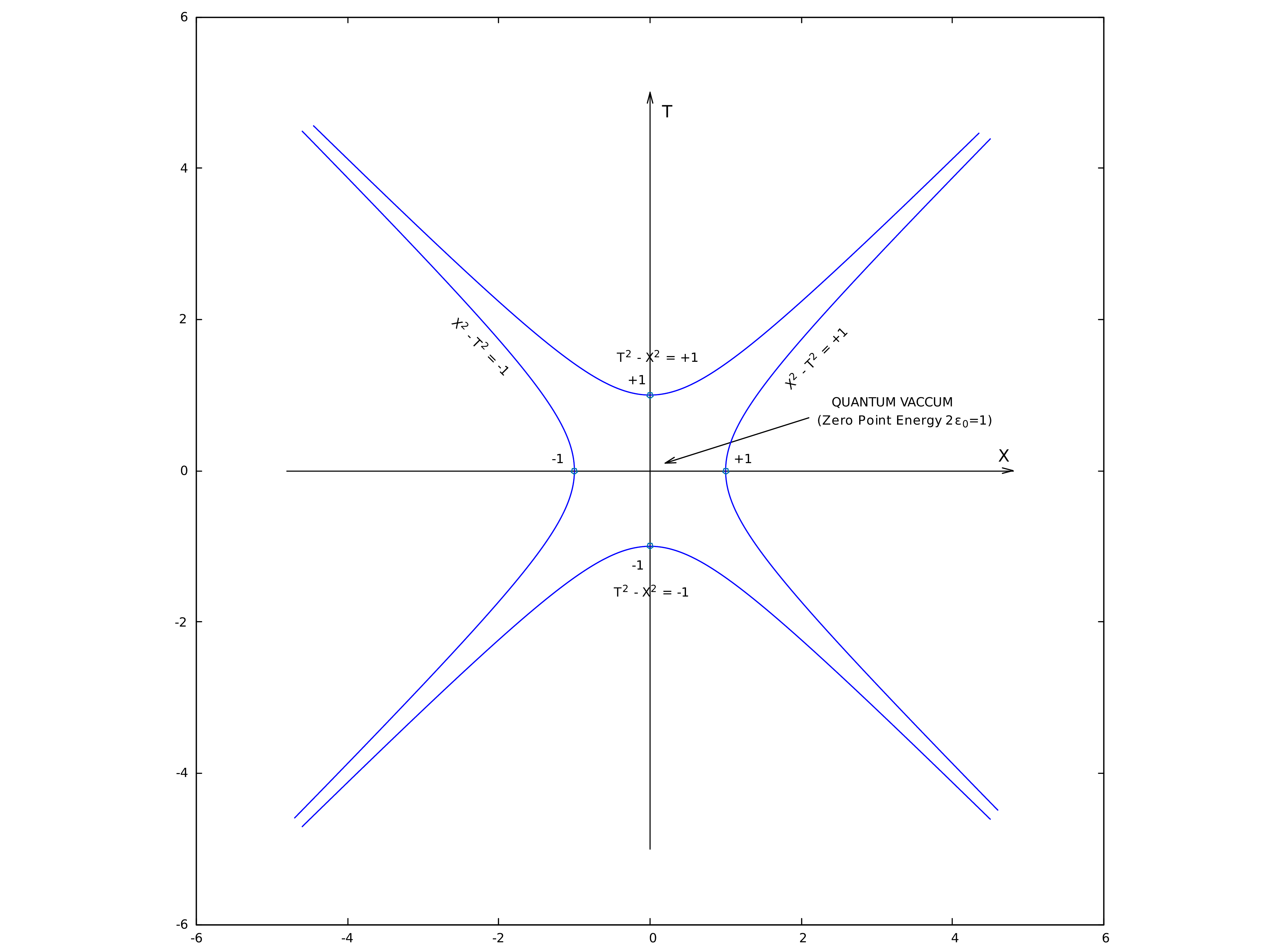}
\caption{{\bf The quantum light cone} (in units of the Planck length). It is generated 
by the {\it quantum hyperbolae} $T^2-X^2 = \pm [X,T] = \pm 1$. 
For comparison, the classical limit: light cone generators $X =\pm T$, is shown in 
Fig. 2. A {\it new} quantum region does appear inside the four Planck scale hyperbolae:
 The Planck scale vacuum due to the zero-point energy $2 
\epsilon_0 = 1$. The four causal regions dissapear inside this Planck scale 
region. The classical conical  vertex $X = \pm T= 0$  {\it spreaded, smeared or erased} at the 
quantum level. This is  due to the non-zero quantum commutator $[X,T]$
or $\Delta X\Delta T$ uncertainty Eqs (2.7).}
\label{fig2b}
\end{figure}
 
The quantization condition Eq.(2.8) yields in this context the {\it quantum 
levels} of the space-time. The space-time hyperbolic structure is discretized 
in odd number levels, Fig 4. It yields for the global coordinates:
$$
X_n = \sqrt{(2n + 1)} \quad \mbox{ for  all $n$ = 0, 1, 2, ... }
$$
\be
X_n \;_{\;\;n >>1}\; = \sqrt{2\;n}\; + \frac{1}{2\sqrt{2\;n}} + O(1/n^{3/2}),  
\qquad \mbox{large $n$}
\ee
\be
 X_n = 1 + n + O(n^2), \qquad \mbox{low $n$} 
\ee
In terms of the local coordinates $ x $ Eq.(1.4), it translates into the  
quantization:
\be
x_{n \pm} =  [\; X_n \pm \sqrt{ X_n^{2} - 1 }\;] = 
 [\;\sqrt{ 2n +1 } \pm \sqrt{2n}\;]
\ee
The condition $ X_n^2 \geq 1$ simply implies $ n\geq 0$: The 
$ n = 0 $ value corresponds to the Planck scale $(X_0 = 1)$ :
\be
x_{0+} = x_{0-} = 1, \;\; n=0: \qquad \mbox{Planck scale}
\ee
\be x_{n\pm} =  1 \pm \sqrt{2\;n}\; + n +  O (n^2), \qquad 
\mbox{low $n$}
\ee
\be
x_{n+} = 2\sqrt{2\;n}\; - 
\;\frac{1}{2\sqrt{2\;n}} + O (1/n^{3/2}), 
\quad  x_{n-} = \frac{1}{2 \sqrt{2\; n}} + O (1/n^{3/2}), \quad 
 \mbox{large $n$}
\ee

Similar analysis holds for $T_n$ and the inverse local coordinates $t_{n\pm}$:
\be
t_{n\pm} =  [\; T_n \pm \sqrt{T_n^{2} + 1 }\;] = 
\sqrt{2} \; [\sqrt{2n+1} \pm \sqrt{(2n+1) + 1/2}\;]
\ee
In the time-like regions, $X_n$ and $T_n$ are exchanged, 
thus covering the global quantum hyperbolic structure, as shown in Fig.4.

\medskip

A {\bf coherent picture} emerges: 
\begin{itemize}
\item {The large modes $n$ correspond to the semiclassical or 
classical states tending towards the classical continum space-time in the very large $ n $ 
limit} 
\item {The low $n$ are quantum, with the lowest mode 
corresponding to the Planck scale $X_0 = 1$, $x_{0+} = x_{0} = 1$.}
\item {The two $ x_{n\pm}$ 
values indicate the two different and dual ways of reaching the Planck scale: from the 
classical/semiclassical side $x_{n+} >> 1$: the $(+)$ branch, and from the 
quantum $0 \leq x_n \leq 1$ side: the $(-)$ branch. 
The large and low $n$ behaviours precisely account for these two dual classical-quantum domains.}
\end {itemize} 

\medskip

\begin{figure}
\includegraphics[height=14.cm,width=16.cm]{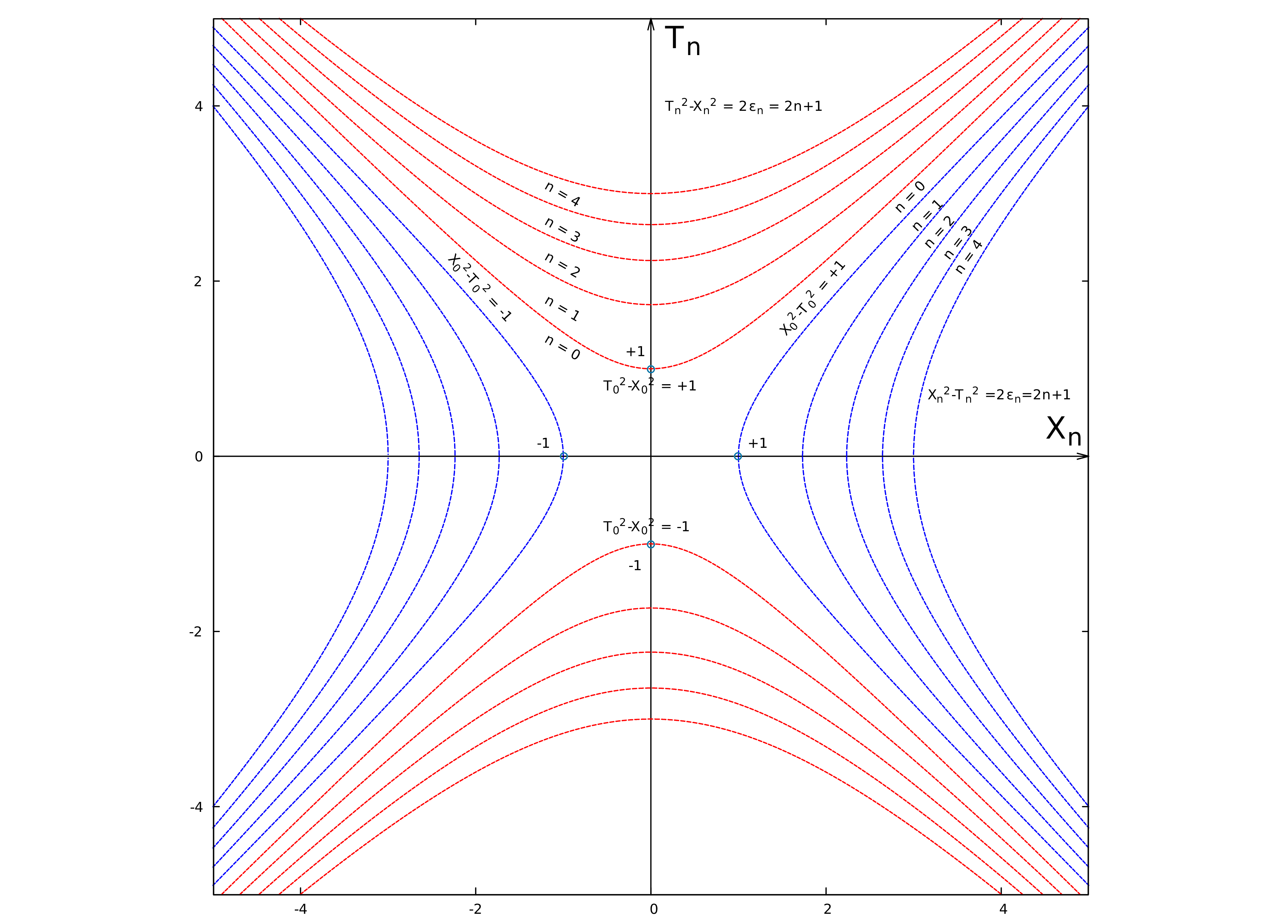}
\caption{ {\bf The quantum space-time and its hyperbolic structure}. It turns out 
to be {\bf discretized} in quantum  hyperbolic levels of odd numbers (in units of the Planck 
length): $X_n^2 - T_n^2 = \pm (2n+1)$ (space-like regions), [$T_n^2 - X_n^2 = \pm (2n +1)$ in the timelike 
regions], $n = 0, 1,2, ...$, $n=0$ being the Planck scale (zero point quantum energy). The $n=0$ quantum hyperbolae generate the {\it quantum light cone}, Fig. 3.
 Low $n$ levels are quantum and bent, large $n$ are classical, less bent tending 
 asymptotically to a classical continum space-time.
 For comparison, the classical space time is shown in Fig.2}
\label{fig3}
\end{figure}

We see that in order to gain physical insight in the quantum Minkowski space-time structure, we can just consider quantum non-commutative coordinates $(X,T)$ with usual quantum commutator $[X,T] = 1$, (1 is here $l_P^2$), and all other commutators zero. In light-cone coordinates 
$$ U = \; \frac {1}{\sqrt{2}} \;(X - T), \qquad V = \; \frac {1}{\sqrt{2}} \;(X + T),$$ 
the quadratic form  (symmetric order of operators) 
\be 
s^2  = [UV + VU] = X^2 - T^2 = (2 VU + 1),
\ee 
determines the relevant component of the quantum distance. This corresponds exactly to the analytic continuation of the euclidean operator $ 2 H = (aa^+ + a^+a)$. The quantum coordinates $(U, V)$ for hyperbolic space-time are the hyperbolic $(T = iP)$  operators ($a, a^+$)  of euclidean phase space and
$VU \equiv N$  is the Number operator. The expectation value
$ (s^2)_n = (2n + 1)$ has as minimal value: $(s^2)^{1/2}_{n=0} = \pm 1$.
Consistently, in quantum space-time we have:
$$
(T^2 - X^2) - 1 \geq 0 : \;\mbox {{\it timelike}}
$$
$$
(X^2 - T^2) - 1 \geq 0:  \;\mbox {{\it spacelike}}
$$
$$
(T^2 - X^2) - (\pm 1) = 0, \;\mbox {{\it null}, (the {\it quantum light-cone)}}.
$$
This is so because only outside the null hyperbolae, ie outside the Planck vaccum region
such notions as distance, and timelike and spacelike signatures can have a meaning, 
Figs 1, 2.

Here we quantized the $(X, T)$ dimensions  which are relevant 
to the light-cone space-time structure.
The remaining spatial transverse dimensions  $ X_{\bot} $ are considered here 
as commuting coordinates, ie having all their commutators zero. For instance, 
in quantum Minkowski space-time:
\be
s^2 = (X^2 - T^2 + X_{\bot j} X_{\bot}^j), \qquad {\bot j} = 2, ...(D-2)
\ee
\be
[X_{\bot j},X] = 0 = [X_{\bot j},T], \qquad [X_{\bot i}, X_{\bot j}] = 0, 
\qquad [P_{\bot i}, P_{\bot j}] = 0
\ee
for all ${\bot}~ i,j = 1, ....., (D-2)$.  $D$ being the total space-time dimensions. In 
particular $D=4$ in the cases considered here.

\medskip

This corresponds to quantize the two-dimensional surface $(X,T)$ relevant for the light-cone structure, 
leaving the transverse spatial dimensions ${\bot}$ with zero commutators. 
This is enough for considering the 
new structure arising in the {\it quantum light cone} and in the 
quantum Rindler and quantum Schwarzschild-Kruskal space-times, for which
 as it is known, the relevant dimensions for the space-time structure are $(X,T)$, 
(and  $x*, t*$) and not the transverse spatial ${\bot}$ dimensions.

This is like one harmonic oscillator in the light cone surface $(X,T)$, and 
no oscillator in the transverse spatial dimensions ${\bot}$.
(Although the $X_{\bot j}$ variables have zero commutators, they could fluctuate). 

\medskip

Here we focus on the space-time quantum 
structure arising from the  relevant non-zero conmutator $[X, T]$ and the {\it quantum 
light cone}. Thus, to follow on the same line of argument, we will consider below the 
quantum Rindler and the quantum Schwarzschild-Kruskal space-time structures. 
Other quantum manifolds where the transverse space $X_{\bot}$ coordinates are 
 also non-commuting can be considered.

\section {Quantum Rindler-Minkowski Space-Time}
 
The above quantum description is still more illustrative by considering the 
transformation:																			                                                              \be X  = \exp{(\kappa x*)} \;\cos(\kappa p*), \qquad
P  = \exp{(\kappa x*)} \; \sin(\kappa p*)
\ee
which is the Rindler phase space representation $(x*, p*)$ of the complete Minkowski 
phase space $(X,P)$. The parameter $\kappa$ is  the dimensionless  (in Planck units)
acceleration. (Here we can express $\kappa = l_P/l = l_P\omega$).  
For classical, ie. non-quantum coordinates  $(X,P)$ we have:
\be
(X^2 + P^2) = \exp{(2\kappa x*)} = 2 \;H, \qquad
(X^2 - P^2) = \exp{(2\kappa x*)}\; \cos(2\kappa p*) 
\ee
We promote now $(X,P)$ to be quantum non-commuting operators, as well as $ (x*, p*)$. 
We get:
\be
(X^2 + P^2) = \exp{(2\kappa x*)} \; \cos (\kappa [x* ,p*])  
\ee
\be
(X^2 - P^2) = \exp{(2\kappa x*)} \; \cos (2\kappa p*) 
\ee
\be
[X,P] = \exp{(2\kappa x*)} \;\sin (\kappa [x*,p*]) , 
\ee 
where we used the usual exponential operator product:\\
$\exp({A})\exp ({B}) = \exp({B})\exp({A})\exp ({[A, B]})$. 

Eqs (3.3)-(3.5) describe the quantum Rindler 
phase space structure. The quantum Rindler space-time follows upon the identification 
$P= iT, p* = it*$: 
$$
X  = \exp{(\kappa x*)} \;\cosh(\kappa t*), \qquad
T  = \exp{(\kappa x*)} \; \sinh(\kappa t*)
$$
\be
(X^2 - T^2) = \exp{(2\kappa x*)} \;\cosh (\kappa [x*,t*]) 
\ee
\be
(X^2 + T^2) = \exp{(2\kappa x*)} \; \cosh (2\kappa t*) 
\ee
\be
[X,T] = \exp{(2\kappa x*)}\; \sinh (\kappa [x*,t*]) 
\ee

\begin{itemize}
\item{We see the new terms appearing due to the quantum conmutators $[X, T]$ and $[x*,t*]$. 
At the classical level: $[X, T] = 0$, $[x*, t*] = 0$ and the known classical 
Rindler-Minkowski equations are recovered.} 

\item{$(X, T)$ and $(x*,t*)$ are quantum coordinates and Eqs (3.6)-(3.8) reveal 
the quantum structure of the Rindler-Minkowski space-time, their classical, semiclassical 
and quantum regions and the classical-quantum 
duality between them. Eqs (3.6) and (3.8) yield:
\be
(X^2 - T^2) = \pm \sqrt{\exp{(4\kappa x*)} + [X,T]^2 } \ee}
\item {We see the role played by the quantum non-zero commutators. 
Also, if the commutators would not be c-numbers, the r.h.s. of
Eqs (3.6)-(3.8) would be just the first terms of the exponential operator expansions, 
but this does not affect the general conclusions here. From Eqs (3.6)-(3.8), 
expectations values and quantum dispersions can be obtained.}  

\item{Eq (3.9) quantum generalize the classical space-time Rindler
"trajectories":  
\be
(X^2 - T^2)_{classical} = \exp{(2\kappa x*)} , \qquad  [X,T] = 0 \; 
\mbox {\it{classically}}
\ee
The quantum analogue of the trajectories ($x*$ = constant) are bendt by the 
non-zero commutator (quantum uncertainty or quantum width) as well as 
the generating Rindler's light-cone.  The classical Rindler's horizons $(x* = -\infty)$ 
 $ X= \pm T$ are quantum mechanically {\it erased},  replaced by
\be
(X^2 - T^2) = \pm \; [X,T] = \pm\;1 :  \;\;\; \mbox {{\it quantum}  Planck scale 
hyperbolae,}
\ee
which are the {\it quantum} "light cone".
At the quantum level, the classical null generators $ X = \pm T$ {\it spread and disappear} 
near and inside the quantum Planck scale vacuum region Eqs (2.9), Fig (3)} 
\item{The quantum algebra Eqs (3.6)-(3.8) and the quantum dispersions and fluctuations 
imply that the four space-time regions (classically I, II, III, IV), are {\it spreaded or 
"fuzzy"}, entangled or {\it erased} at the quantum level, near and inside the Planck 
domain delimitated by the four Planck scale hyperbolae Eq (3.11), Figs 3 and 4.}
\item{Fig 4 shows the quantum {\it discrete levels} of Minkowski-Rindler space-time and 
all the previous discussion applies here
\be X_n^2 - T_n^2 = \pm (2n+1), \qquad n = 0,1,2,...\ee
"Exterior" Rindler regions to the Planck scale hyperbolaes $ (X^2-T^2)_{n=0}$ = $\pm 1$ 
contain the quantum, semiclassical and classical behaviours, from $n = 0$ and the 
low $ n$ to the large ones, 
which became more classical and less bendt, in agreement with the 
classical-quantum duality of space-time structure.} 
\item{The interior region to the $n=0$ levels is the full quantum Planck scale domain. 
The "future" and "past" regions are composed by levels from the very quantum (the Planck $n=0$ 
hyperbolae and low $n$), to the semiclassical and classical  (large $n$) levels  
$(X_n, T_n)$.}
\item{The Rindler levels  $(x*_{n\pm}, t*_{n\pm})$ 
 follow from Eqs (2.13)-(2.17) for $(x_{n\pm}, t_{n\pm})$: \be
x_{n\pm} = \exp{(\kappa x*_{n\pm})} =  [\; X_n \pm \sqrt{ X_n^{2} - 1 }\;] = 
[\;\sqrt{ 2n +1 } \pm \sqrt{2n}\;]
\ee \be
t_{n\pm} = \exp{(\kappa t*_{n\pm})} =  [\; T_n \pm \sqrt{T_n^{2} + 1 }\;] =
[\sqrt{2n+1} \pm \sqrt{(2n+1) + 1/2}\;] 
\ee} 
\item{Due to the quantum space-time width, quantum light-cone or  
quantum dispersion and fluctuations, and the quantum Planck scale
nature of the interior region, the difference between the four causal 
regions I, II, III, IV is {\it quantum 
mechanically erased} in the Planck scale region.
The classical copies or halves (I, II) and (III, IV) became  
{\it one only quantum world}.}
\item{This provides further support to the {\it antipodal identification} 
of the two  space-time copies which are classically or semiclassically the space and time 
reflections of each other and which are classical-quantum duals
of each other, and therefore supports the antipodally symmetric quantum theory, 
refs [3], [4], [5], [6].
The classical/semiclassical antipodal space-time symmetry and the CPT symmetry belong  to the general QG classical-quantum duality symmetry ref [1].}
\end {itemize}

\section{Quantum Schwarzschild-Kruskal Space-Time}
 
Let us now go beyond the classical Schwarzschild-Kruskal space-time structure
and extend to it the findings of the sections II, III 
above.                                                                                                                         
\medskip

We have seen in ref [1] that in the complete analytic extension or global structure 
of the Kruskal space-time underlies a classical-quantum duality structure: 
The external or visible region and its mirror copy  
are the classical or semiclassical gravitational domains while the internal region 
is fully quantum gravitational -Planck scale- domain.
A duality symmetry between the two external regions, and between the internal and 
external parts {\it shows up} as a {\it classical - quantum duality}. External and internal 
regions meaning now with respect to the hyperbolae $X^2 - T^2 = \pm1$.

\medskip

In order to go beyond the classical - quantum dual structure 
of the Schwarzschild-Kruskal space-time and to account 
for a {\it quantum}  Schwarzshild-Kruskal description of space-time,
we proceed as with the quantum Minkowski-Rindler 
space-time variables in previous section.
The phase space and space-time coordinate transformations are the same in both 
Rindler and Schwarzschild cases. The classical Kruskal 
phase space coordinates $(X,P)$ in terms of the Schwarzschild 
phase-space representation $(x*, p*)$  are given by
\be 
X =  \exp{(\kappa x*)}\cos (\kappa p*), \qquad \;
P =  \exp{(\kappa x*)}\sin (\kappa p*)
\ee
\be
(X^2 + P^2) = \exp{(2\kappa x*)} \;  = 2 \;H, \qquad 
(X^2 - P^2) = \exp{(2\kappa x*)} \cos (2 \kappa p*)
\ee
with the Schwarzschild star coordinate $x*$:
\be 
\qquad \exp(\kappa x*) = \sqrt{2\kappa r - 1} \;\exp(\kappa r), \quad 2\kappa r > 1
\ee
being $ \kappa $ the dimensionless (in Planck units) gravity acceleration or surface gravity.
Another patch similar to Eqs (4.1)-(4.3) but with $X$ and $P$ exchanged and $x*$ defined by
$ \exp(\kappa x*) = \sqrt{1- 2\kappa r}\;\exp(\kappa r) $, holds for $2 \kappa r < 1$.

\medskip

By promoting  $(X, P)$ to be quantum coordinates, ie non-commuting
operators, and similarly for $(x*, p*)$, 
yields Eqs (3.3)-(3.5). They provide in this case the quantum Kruskal's 
phase space coordinates $(X,P)$ in terms of the quantum Schwarzschild coordinates 
$(x*, p*)$ with $x*$ given by Eq. (4.3). The corresponding quantum Kruskal's 
space-time follow upon the identification: 
$ P = iT, \; p* = it* $. In terms of Schwarzschild's space-time coordinates $(x*, t*)$ it yields:
\be 
X =  \exp{(\kappa x*)}\cosh (\kappa t*), \qquad \;
T =  \exp{(\kappa x*)}\sinh (\kappa t*)
\ee
\be
(X^2 - T^2) = \exp{(2\kappa x*)} \; \cosh (\kappa [x*,t*]) 
\ee
\be
(X^2 + T^2) = \exp{(2\kappa x*)} \; \cosh (2\kappa t*) 
\ee
\be
[X,T] = \exp{(2\kappa x*)}\; \sinh (\kappa [x*,t*])  
\ee

We see the new terms appearing due to the quantum conmutators. At the classical level: 
$$ [X, T] = 0,  \qquad [x*, t*] = 0 \quad (\mbox{classically})$$ and  
the known classical Schwarzschild-Kruskal equations are recovered. 

Eqs (4.5)-(4.7) describe the quantum Schwarzschild-Kruskal space-time structure and its 
properties, we analyze them below. 
Upon the identification $P=iT$, the quantum Kruskal ligth-cone variables 
\be U = \frac{1}{\sqrt {2}} (X - T), \quad V = \frac{1}{\sqrt {2}}(X + T) \ee
 in hyperbolic space are the $(a, a^+)$ operators Eqs (2.4). 
The quadratic form (symmetric order of operators):
$$ 2H = UV + VU = X^2 - T^2 = (2UV+ 1), 
\qquad UV = N \equiv \mbox {number operator}, $$
yields the quantum hyperbolic structure and the discrete hyperbolic space-time levels:
\be
X_n^2 - T_n^2 = (2n + 1) \quad \mbox{and} \quad  T_n^2 - X_n^2 = (2n + 1), 
\qquad (n= 0, 1, ...) 
\ee
The amplitudes $(X_n, T_n)$  are  $ \sqrt {2n+1} $ and 
follow the same Eqs (2.10)-(2.12) and Fig 4. We describe the quantum structure below.

\subsection{IV. No horizon, no space-time singularity and only one Kruskal 
world}

From Eqs (4.5)-(4.7), expectation values and quantum dispersions 
can be obtained.  For instance,   
the equation for the quantum hyperbolic "trajectories" is
\be
(X^2 - T^2) = \pm \sqrt{\exp{(4\kappa x*)} + [X,T]^2 } =  
\pm \sqrt{ (1 - 2 \kappa r)^2\exp{(4\kappa r)} + [X,T]^2 }
\ee
The characteristic lines and what classically were the light-cone generating 
horizons $X = \pm T$ (at $2 \kappa r = 1$, or $x* = - \infty$) are now: 
\be
X = \pm \sqrt{ \;T^2 + [X,T]^2\;} \quad \mbox {at $2\kappa r= 1$:} \quad X \neq \pm T \; ,
\; \mbox{{\it no horizons} }
\ee
We see that $X \neq \pm T$ at $2\kappa r= 1$ and the null horizons are {\it erased}.

Similarly, in the interior regions the classical hyperbolae $(T^2 - X^2)_{classical} = \pm 1$ 
which described the known past and future classical singularity $ r = 0, (x* = 0)$ 
are now replaced by :
$$
(T^2 - X^2) = \pm \sqrt{\; 1 + [X,T]^2 \;} = \pm \sqrt{2}\quad \mbox {at $r= 0$:}\quad (T^2 - X^2) \neq \pm 1 \; 
 \mbox{{\it no singularity}} 
$$ 
\be
(T^2 - X^2)_{classical} = \pm 1 \quad \mbox {at $r= 0$ \; {\it classically}}
\ee
The classical singularity $ r= 0 = x* $ is  \textit{quantum mechanically
smeared or erased} which is what is expected in a quantum space-time
description. 

\begin{itemize}

\item{The right and left "exterior" regions to the quantum Planck hyperbolae \\
$(X^2 -T^2)_{n=0} = \pm 1$ in Fig. 4 contain all quantum, semiclassical and classical 
allowed levels from the $n=0$ (Planck scale),  low $n$ 
(quantum) to the intermediate and large $n$ (classical) behaviours.}

\item{Similarly, the future and past regions to the quantum Planck hyperbolae\\
$(T^2 - X^2)_{n=0} = \pm 1$, contain {\it all} allowed levels and behaviours. 
There is {\it not} $r = 0 = x*$ singularity boundary in the quantum space-time.}

\item{$(T^2 - X^2)_{n=0} = \pm 1$ are the quantum -Planck scale- hyperbolae which replace the 
classical null horizons $(X = \pm T)_{classical}$ at $x* = -\infty, 2\kappa r = 1$ 
in the quantum space-time.}

\item{$(T^2 - X^2) = \pm \sqrt {2}$ are the quantum hyperbolae which 
replace the classical singularity:  $(T^2 - X^2)_{classical} (r=0) = \pm 1$. 
Moreover,
the quantum hyperbolae $(T^2 - X^2) = \pm \sqrt {2}$  lie {\it outside} the 
allowed quantum hyperbolic levels. 
They do not correspond to any of the allowed quantum levels Eqs (4.10), $n = 0, 1, 2, ...$
and therefore, they {\it are excluded} at the quantum level: The
singularity is {\it removed} out from the quantum space-time.} 

\item{There are {\it no} singularity boundaries at $(T^2 - X^2)(2 \kappa r=1) = \pm 1 $  
nor at $(T^2 - X^2) = \pm \sqrt {2}$ at the quantum level. 
The quantum space-time {\it extends} without boundary beyond the Planck hyperbolae
$(T^2 - X^2)(n=0)  = \pm 1 $ towards {\it all} levels: from the more quantum (low $n$)
levels to the classical (large $n$) ones, as shown in Fig.4.}

\item{The internal region to the four quantum Planck hyperbolae $(T^2 - X^2)(n=0)  = \pm 1 $ is 
{\it totally} quantum and within the Planck scale: this is the quantum vacuum or "zero 
point Planck energy" region. This confirms and expands our result in ref [1] about the 
 {\it quantum interior} region of the black hole.}
 
\item{The null horizons  {\it disappeared} at the quantum level.
Due to the quantum $[X, T]$ commutator, quantum $(X, T)$ dispersions and fluctuations,
the difference between the four classical Kruskal regions (I, II, III, IV)  
 {\it dissapears} in the Planck scale domain. 

This provides further support to the {\it antipodal 
identification} of the two Kruskal copies which are classically and semiclassically 
the space-time reflection of each other, and which translates into the 
CPT symmetry and antipodally symmetric states refs [3],[4],[5],[6].}

\item {The levels in terms of the Schwarzschild variables $(x*_{n\pm}, t*_{n\pm})$ follow from  
Eqs (3.13), (3.14) for $(x_{n\pm}, t_{n\pm})$, being 
 $x = \exp {(\kappa x*)}$, and $t = \exp {(\kappa t*)}$:
\be
x_{n\pm} = [\; \sqrt {2 \kappa r_{n\pm} - 1}\;]\; \exp{(\kappa r_{n\pm})} 
= [\;\sqrt{ 2n +1 } \pm \sqrt{2n}\;]    
\ee 
\be
t_{n\pm} =  [\;\sqrt{2n+1} \pm \sqrt{(2n+1) + 1/2}\;],
\ee
which complete all the levels. Their large $n$ and low $n$ behaviours 
follow Eqs (2.14)-(2.16) and  their respective clasical-quantum duality properties.}
\end{itemize}

\section{Mass quantization. The whole mass spectrum}

\medskip

$(X_n,T_n) , (x_n, t_n) $ are given in Planck (length and time) units.  In terms of the mass 
global variables  $ X = M/m_P$, or the local ones $x = m/m_P$, Eqs (1.4), (1.7),
it translates into the mass levels:
\be
M_n = m_P \sqrt{(2n + 1)},\qquad \mbox{all} \;\;n = 0, 1, 2, ....
\ee
\be
M_n \; _ {n >>1}\;= 
\; m_P \; [\sqrt{2\;n}\; + \frac{1}{2\sqrt{2\;n}} + O(1/n^{3/2})\;],  
\ee
\be
m_{n \pm} =  [\; M_n \pm \sqrt{ M_n^{2} - m_P^2 }\;], 
\ee
 The condition $M_n^2 \geq m_P^2$ simply corresponds to the whole spectrum $n\geq 0$:
\be
m_{n \pm} = m_P \; [\;\sqrt{ 2n +1 } \pm \sqrt{2n}\;]
\ee
\be
m_{0 +} = m_{0 -} = M_0 = m_P ,\qquad n=0: \; \mbox {Planck mass $m_P$}  
\ee
\be m_{n +} = m_P \; [\;2\sqrt{2\;n}\; - \;\frac{1}{2\sqrt{2 n}} 
+ O (1/n^{3/2})\;],  \quad \mbox{large $n$ \; $=> \;\; m_+$ larger than $m_P$} 
\ee
\be
m_{n-} = \frac{m_P}{ 2 \sqrt{2\; n}} + O(1/n^{3/2}),
\quad \mbox{large $n$ \; $=> \;\; m_-$ smaller than $m_P$} 
\ee
\begin{itemize}
\item {The mass quantization here holds for {\it all masses}, not only for black holes. 
Namely, the quantum mass levels are associated to the quantum space-time structure. 
Space-time can be parametrized by {\it masses} ("mass coordinates"),  
just related to length and time, as the QG variables, 
on the same footing as space and time variables.
In Planck units, any of these variables (or another convenient set) can be used.}

\item{The two ($\pm$) {\it dual mass} branches (classical and quantum) 
Eqs (5.4)-(5.7) correspond to the large and small 
masses with respect to the Planck mass $m_P$, they cover the {\it whole} mass range:  
From the Planck mass: branch (+), and from zero mass till near 
the Planck mass: branch (-).}  

\item {As $n$ increases, masses in the branch (+) {\it increase} 
from $m_P$ covering  all the mass spectrum of gravitational
objects till the largest masses.  Masses are quantized 
as $ m_P (2\sqrt {2n}) $ as the dominant term, Eq (5.6). For very large $n$ 
the spectrum becames continuum. Macroscopic objects and astronomical masses 
belong to this branch (gravitationnal branch).} 

\item {As $n$ increases, masses in the branch (-) {\it decrease}:
The branch (-) covers the masses {\it smaller} than $m_P$ from the zero mass 
to masses remaining smaller than the Planck mass: large $n$ behaviour of branch (-) Eq.(5.7). 
The quantum elementary particle masses belong to this branch (quantum particle branch).} 

\item{Black hole masses belong to both branches (+) and (-). 
Branch (+) covers all macroscopic and astrophysical black holes as well as semiclassical 
black hole quantization $\sqrt{n}$  till masses nearby the Planck mass.}

\item{The microscopic black holes, {\it quantum} black holes (with masses 
near the Planck mass and smaller 
 till the zero mass, ie as a consequence of black hole evaporation), belong to the branch (-). 
The  branches (+) and (-) cover {\it all} the black hole
masses. The black hole masses in the process of black hole evaporation go 
from branches (+) to (-). Black hole ends its evaporation in branch (-) decaying as a 
pure quantum state.} 

\item{Black hole evaporation  
is not the subject of this paper but our results  here have implications for it.
The last stage of black hole evaporation
and its quantum decay belong to the quantum branch (-).
Black hole evaporation is thermal (ie a mixed state) in its semiclassical gravity phase 
(Hawking radiation) and it is non thermal in its last quantum stage (pure quantum decay)
 refs [2], [7], [8].
In its last phase (mass smaller than the Planck mass $m_P$), the state 
is not anymore a black hole, but a pure (non mixed) quantum state, 
decaying like a quantum heavy particle.}
\end {itemize}

\section{Conclusions}

\begin {itemize}

\item{We have investigated here the quantum space-time structure 
arising from the relevant non-zero space-time commutator $[X, T]$, 
or non-zero quantum uncertainty $ \Delta X \Delta T$ by considering 
{\it quantum} coordinates $(X,T)$. 
The remaining transverse spatial coordinates $X_{\bot}$ 
have all their commutators zero. This is enough to capture the essential features of the 
new quantum space-time structure. }

\item{We found  the {\it quantum light cone}: 
It is generated by the quantum Planck hyperbolae 
$ X^2 - T^2 = \pm [X, T]$ due to the
quantum uncertainty $[X,T]= 1 $
They replace the classical light cone generators $ X = \pm T $ which are 
{\it quantum mechanically
erased}. Inside the four Planck hyperbolae there is a enterely 
{\it new quantum region} within the Planck
scale and below which is a purely quantum vacuum or zero-point Planck energy region}

\item {The quantum non-commuting coordinates $(X,T)$ 
and the transverse commuting spatial coordinates $X_{\bot j}$
generate the quantum two-sheet hyperboloid $ X^2 - T^2 + X_{\bot j} X_{\bot}^j = \pm 1$.} 

 \item {We found the quantum Rindler and the quantum Schwarzschild-Kruskal
space-time structures: we considered the relevant quantum non-commutative 
coordinates and the quantum 
hyperbolic "light cone" hyperbolae. They generalize the classical 
known Schwarzschild-Kruskal 
structures and yield them in the classical case (zero quantum commutators). 
At the quantum level, the classical null horizons $X = \pm \; T$ 
are {\it erased}, and the $r=0$ classical singularity {\it dissapears}. 
Interestingly enough, the Kruskal space-time structure turns out to be {\it discretized}
in {\it quantum hyperbolic levels} $(X_n^2 - T_n^2) = \pm (2n+1), \, n = 0,1,2...$.
Moreover, the $ r= 0$ singular -hyperbola is quantum mechanically {\it excluded}, it does 
{\it not} belong to any of the quantum {\it allowed} levels.}

\item {The quantum Schwarzschild-Kruskal space-time {\it extends without boundary 
and without any singularity} in quantum discrete allowed levels beyond  the quantum Planck 
hyperbolae $X_0^2 - T_0^2 = \pm 1$, from the Planck scale 
$(n=0)$ and the very quantum levels (low $n$) to the quasi-classical and classical levels (intermediate and large $n$),
and asymptotically tend to a continuum classical space-time for very large $n$.}  

\item{The quantum mass levels here hold for {\it all} masses. The two ($\pm$) 
{\it dual mass} branches correspond to the larger and smaller masses 
with respect to the Planck mass $m_P$ respectively, they cover the 
{\it whole mass range} from the Planck mass in branch (+) untill the 
largest astronomical masses, and from zero mass in branch (-) 
in the elementary particle domain till near the Planck mass.
As $n$ increases, masses in the branch (+) {\it increase} (as $2 \sqrt{2n}$).
For very large $n$ the spectrum becames continuum. 
Masses in the branch (-) {\it decrease} in the large $n$ behaviour, precisely as 
$ 1 /(2 \sqrt{2n}) $, the dual of branch (+). 
The whole mass levels are provided in Section V above. 
Black hole masses belong to both branches (+) and (-).}

\item{The quantum end of black hole evaporation is not the central 
issue of this paper, but our results here have consequences for this problem which 
we will discuss elsewhere: The quantum black hole decays into elementary particle states, 
that is to say pure (non mixed) quantum 
states, in discrete levels and other implications ref [9].}

\item{ We can similarly think in quantum string coordinates (collection of point oscillators) 
to describe the quantum space-time structure,
(which is {\it different} from strings propagating on a fixed space-time background). 
This yields similar results to the results found here with a quantum {\it 
hyperbolic space-time width} and hyperbolic structure for the 
characteristic lines and light cone generators, or for the 
space-time horizons ref [9].

Moreover, we see that the mass quantization $m_P\sqrt {n}$ 
we found here, ie Eq (5.1), Eq (5.4), is like the string mass quantization 
$M_n = m_s  \sqrt{n}$,  $n = 0, 1, ...$
with the Planck mass $m_P$ instead of the fundamental string mass 
$m_s$,  ie $G/c^2$ instead of the string constant $\alpha'$.}

\item {Here we focused on the space-time quantum 
structure arising from the relevant non-zero commutator $[X, T]$: the {\it quantum 
light cone} which is relevant for the Minkowski, Rindler and 
the Schwarzschild-Kruskal {\it quantum} space-time structures. 

Quantizing the higher dimensional transverse dimensions $X_{\bot j}$ does not change 
the basic new quantum structure here. In another manifolds, there will be specific 
$(D-2)$ spatial transverse contributions.
Quantum non-commuting transverse coordinates important for another type of manifolds 
will be considered elsewhere, ref [9].}
  
\end{itemize}

\bigskip

{\bf ACKNOWLEDGEMENTS}

\bigskip

The author thanks G.'t Hooft for interesting and stimulating communications on 
several occasions, M. Ramon Medrano for useful discussions and encouredgement and 
F.Sevre for help with the figures. 
The author acknowledges the French National Center of Scientific Research (CNRS)
for Emeritus contract. This work was performed in LERMA-CNRS-Observatoire de Paris-
PSL Research University-Sorbonne Universit\'{e} Pierre et Marie Curie.

\end{document}